# Kilovolt-class Vertical (011) β-$Ga_2O_3$ Schottky Diodes for High Voltage and High Temperature Applications

Emerson J. Hollar,[1] Aaron M. Adams,[2] Nolan S. Hendricks,[2] Andrew J. Green,[2] and Esmat Farzana[1*]

[1]Department of Electrical and Computer Engineering, Iowa State University, Ames, IA 50011, USA

[2]Air Force Research Laboratory, Sensors Directorate, OH 45433, USA

***Abstract***– We report kilovolt-class vertical (011) β-$Ga_2O_3$ Schottky diodes utilizing high-permittivity dielectric $TiO_2/Al_2O_3$ field-plate for high voltage and high temperature applications. A systematic study was performed with Pt/(011) β-$Ga_2O_3$ Schottky barrier diodes (SBDs) with varied diameters from 100 μm to 300 μm, for both with and without field-plate, that revealed excellent consistency of reverse blocking performance regardless of diode area. The field-plate SBDs achieved superior breakdown voltages (3.90 - 4.16 kV) compared to the diodes without field-plate (3.06 - 3.16 kV) owing to effective edge termination. Furthermore, we explored high-temperature performance of the field-plate (011) β-$Ga_2O_3$ diodes that revealed excellent forward conduction properties and high rectification ratio (~$10^9$) throughout the temperature range (25 - 150 ℃). At reverse bias, the field-plate diodes also exhibited kV-range breakdown voltage with no evident increase in leakage current up to the explored elevated temperature of 150 ℃. The low area-dependence of breakdown voltage, excellent forward transport properties, and minimal reverse leakage at both high-voltage and elevated temperature demonstrate the exciting potential of (011) β-$Ga_2O_3$ SBDs in high-temperature power switches. Thus, our work demonstrates the strategy of advancing the performance of vertical β-$Ga_2O_3$ SBDs by utilizing the advantageous (011) β-$Ga_2O_3$ epilayers with low background doping, reduced killer dislocation effects, and effective field management for high-voltage and high-temperature applications.

*E-mail: efarzana@iastate.edu

The ultrawide bandgap (UWBG) β-$Ga_2O_3$ has attained wide-spread appeal for the realization of compact high-power semiconductor devices, supported through a ~4.8 eV band-gap, controlled shallow dopants, and projected high critical breakdown field (~8 MV/cm).[1–3] Additionally, β-$Ga_2O_3$ offers an exceptional advantage of melt-grown native substrates amidst all other wide bandgap (WBG) and UWBG semiconductors, offering the distinct opportunity for low-cost and large-scale production of (U)WBG power devices. Promising device performance has been demonstrated in vertical β-$Ga_2O_3$ power diodes through a variety of design and fabrication techniques including field-plate,[2-9] trench structures,[10-14] guard rings,[15,16] deep etch,[17-20] and ion implantation.[21] However, most of these reported devices utilized halide vapor phase epitaxy (HVPE)-grown (001) orientated β-$Ga_2O_3$ epilayers that suffer from limited drift layer thickness (≤10 μm) and unintentional donor impurities that severely limit the reverse blocking performance for vertical power devices.[22,23] Furthermore, the (001) β-$Ga_2O_3$ epilayer growth has been reported to facilitate killer dislocation defects that create leakage paths across the drift layer of vertical devices and degrade reverse breakdown performance.[3,24,25] Hence, to upscale β-$Ga_2O_3$ power devices towards higher voltage ratings, thicker drift layers with reduced doping and killer dislocation defects remains a fundamental requirement.

To overcome these critical challenges, alternative crystal orientations have been explored in recent years, such as the (011) β-$Ga_2O_3$, which is promising to mitigate the effects of dislocation-assisted leakage in vertical β-$Ga_2O_3$ power devices since the dislocations remain parallel to the (011) β-$Ga_2O_3$ plane.[22,25,26] Furthermore, the (011) β-$Ga_2O_3$ is reported to reduce unintentional [Cl] donor impurity incorporation in HVPE-grown epitaxial wafers that enables lower background doping ($<5\times10^{15}$ cm$^{-3}$) compared to that of conventional (001) β-$Ga_2O_3$ with practical low background doping limit ~$8\times10^{15}$ cm$^{-3}$.[22,23] High-quality (011) β-$Ga_2O_3$ homoepitaxial films have also been reported by other prominent growth methods, such as metalorganic chemical vapor deposition (MOCVD), with promising demonstration of thick epilayer, low controlled doping ability, and smooth surface morphology.[27,28] These advancements have already led to multi-kV class vertical (011) β-$Ga_2O_3$ power device demonstrations, including Schottky diodes,[3,29] heterojunction diodes,[29,30] and transistors.[23]

In addition to the high-voltage application, the UWBG power devices are considered superior candidates for high-temperature and harsh environment electronics due to their advantageous high breakdown field, predicted improved thermal stability, and low intrinsic carrier concentration compared to silicon and even WBG counterparts. However, there has been limited research on the high-temperature breakdown performance of multi-kV class vertical β-$Ga_2O_3$ Schottky barrier diodes (SBDs). Among the existing literature, most reports have focused on temperature-dependent forward bias characteristics and a limited range of reverse bias voltage (≤500 V) of the vertical β-$Ga_2O_3$ SBDs.[7,9,29,31-33] Moreover, due to the early stage of research, the high-temperature performance of the vertical (011) β-$Ga_2O_3$ SBDs at high reverse bias has remained unexplored. Such an investigation is critically important to assess the thermal stability and potential of the (011) β-$Ga_2O_3$ SBDs in kV-scale high-temperature electronics.

To address this gap, we investigated the high-voltage and high-temperature performance of vertical (011) β-$Ga_2O_3$ SBDs fabricated with high-permittivity (κ) dielectric $TiO_2$/$Al_2O_3$ based field-plate. Here, the top thick $TiO_2$ (κ ~43) in the field-plate stack can enable effective field reduction due to its high permittivity properties as demonstrated in our prior work and earlier reports[6,34,35] where the underlying $Al_2O_3$ can suppress leakage at the field-plate interface owing its large positive conduction band offset (CBO) with β-$Ga_2O_3$ (~1.7 eV)[36] compared to that of $TiO_2$ with negative CBO (~0.3 eV).[34] Our fabricated diodes showed excellent consistency of breakdown voltages between 3.9-4.2 kV across the different diameters (100-300 μm) and maintained ultra-low leakage current until breakdown. Furthermore, we examined the high-temperature stability of the vertical (011) β-$Ga_2O_3$ field-plate diodes, up to 150 ℃, where the SBDs exhibited excellent forward transport properties, rectification ratio of ~$10^9$, and kV-range breakdown voltage with minimal leakage throughout the explored temperature range. Hence, our work demonstrates promising capabilities of vertical (011) β-$Ga_2O_3$ SBDs for multi-kV class and high-temperature power electronics.

The vertical SBDs were fabricated on commercially grown HVPE (011) β-$Ga_2O_3$ epilayers with 30 μm nominal thickness and low background doping (~2 × $10^{15}$ $cm^{-3}$), grown on ~400 μm thick Sn-doped (011) β-$Ga_2O_3$ substrate (doping ~5 × $10^{18}$ $cm^{-3}$) as shown in Fig. 1(a). The Ohmic contact was formed using a $BCl_3$-based reactive ion etch (RIE) backside treatment followed by e-beam evaporation of Ti/Au (30 nm/150 nm) metal stack. The Ohmic metal stack was subsequently annealed at 470 °C in nitrogen for 1 minute using rapid thermal annealing (RTA). The Schottky anode contacts were patterned for diodes with 100 μm, 200 μm, and 300 μm diameters, that were deposited using e-beam evaporation of Pt contact. The field-plate dielectric stack was formed by a 200 °C atomic layer deposition (ALD) of thin (~9.7 nm) $Al_2O_3$ followed by sputtered deposition for thick (~160 nm) $TiO_2$ following our prior work.[6] This thin $Al_2O_3$ layer was formed by ALD to protect the field-plate dielectric interface from plasma-induced sputter damage during the subsequent thick $TiO_2$ layer deposition. Afterwards, the field-plate dielectric stack was patterned for via opening using a combination of a dry etch with $BCl_3$-based RIE and wet etch with dilute HF that exposed the Pt Schottky contact. Finally, the field-plate was formed with a sputtered Pt cap/$PtO_x$ stack using a lift-off process where the $PtO_x$ was formed by reactive sputtering of low-power (55W) Pt in the presence of Ar:$O_2$ flux at 10:5 sccm following our previous work.[1,3] The lengths of the field-plate ($L_{FP}$) were 30 μm, 40 μm, and 60 μm for diodes with diameters 100 μm, 200 μm, and 300 μm respectively. Some diodes were kept as baseline SBDs without field-plate in the same sample to enable fundamental electronic characterization of the diode properties as well as evaluate the efficacy of edge termination.

The baseline SBDs without edge termination were characterized first using 1 MHz capacitance voltage (C-V) analysis that reveled doping concentration of ~$2\times10^{15}$ $cm^{-3}$ and a Schottky barrier height (SBH) of 1.59 eV [Fig. 1(b)]. The field-plate diodes were subsequently characterized with forward current density-voltage (J-V) characteristics that revealed excellent transport characteristics with near unity ideality factor across all diode diameter cases, as depicted in Fig. 1(c). This indicates a dominant thermionic emission (TE) transport mechanism at room temperature which led to Schottky barrier height (SBH) from forward J-V analysis as 1.44 eV, 1.49 eV, and 1.51 eV for diodes with diameters 100 μm, 200 μm, and 300 μm, respectively.[3,37] For each diameter case, several diodes were characterized where the mean and standard deviation of the SBH appeared as 1.48 $\pm$0.027 eV, 1.50 $\pm$0.023, and 1.51 $\pm$0.012 eV, indicating excellent uniformity of SBH across the diameter. The minimum differential specific on-resistance ($R_{on,sp}$) of the representative SBDs of Fig. 1(c) was also analyzed from the forward J-V characteristics, and extracted to be 35.99 mΩ-$cm^2$, 41.50 mΩ-$cm^2$, and 46.32 mΩ-$cm^2$ for diodes with diameters 100 μm, 200 μm, and 300 μm, respectively [Fig. 1(d)].

The breakdown voltage ($V_{br}$) of the SBDs were characterized afterwards by applying reverse bias until the diodes experienced destructive failure. To prevent the breakdown of surrounding air during testing, the devices were fully submerged in FC-40 Fluorinert dielectric liquid. The SBDs did not exhibit breakdown up to the 3 kV range of our parametric analyzer at Iowa State University (ISU) under Fluorinert. Hence, the breakdown was completed afterwards by accessing a 10 kV probe station at the Air Force Research Laboratory (AFRL). The SBDs without a field-plate reached a reverse breakdown voltage of 3.16 kV, 3.05 kV, and 3.16 kV, for diodes with diameters 100 μm, 200 μm, and 300 μm, respectively [Fig. 2]. With the incorporation of field-plate, the SBDs exhibited enhanced reverse breakdown performances of 4.16 kV, 4.13 kV, and 3.90 kV with diameters of 100 μm, 200 μm, and 300 μm, respectively. The excellent agreement in breakdown voltages of the vertical (011) β-$Ga_2O_3$ SBDs, despite increased diode area, demonstrates the benefits of the (011) β-$Ga_2O_3$ epilayer devices with reduced killer dislocation defects. This brings enormous potential for high-voltage (011) β-$Ga_2O_3$ toward scaling to large-area power devices.

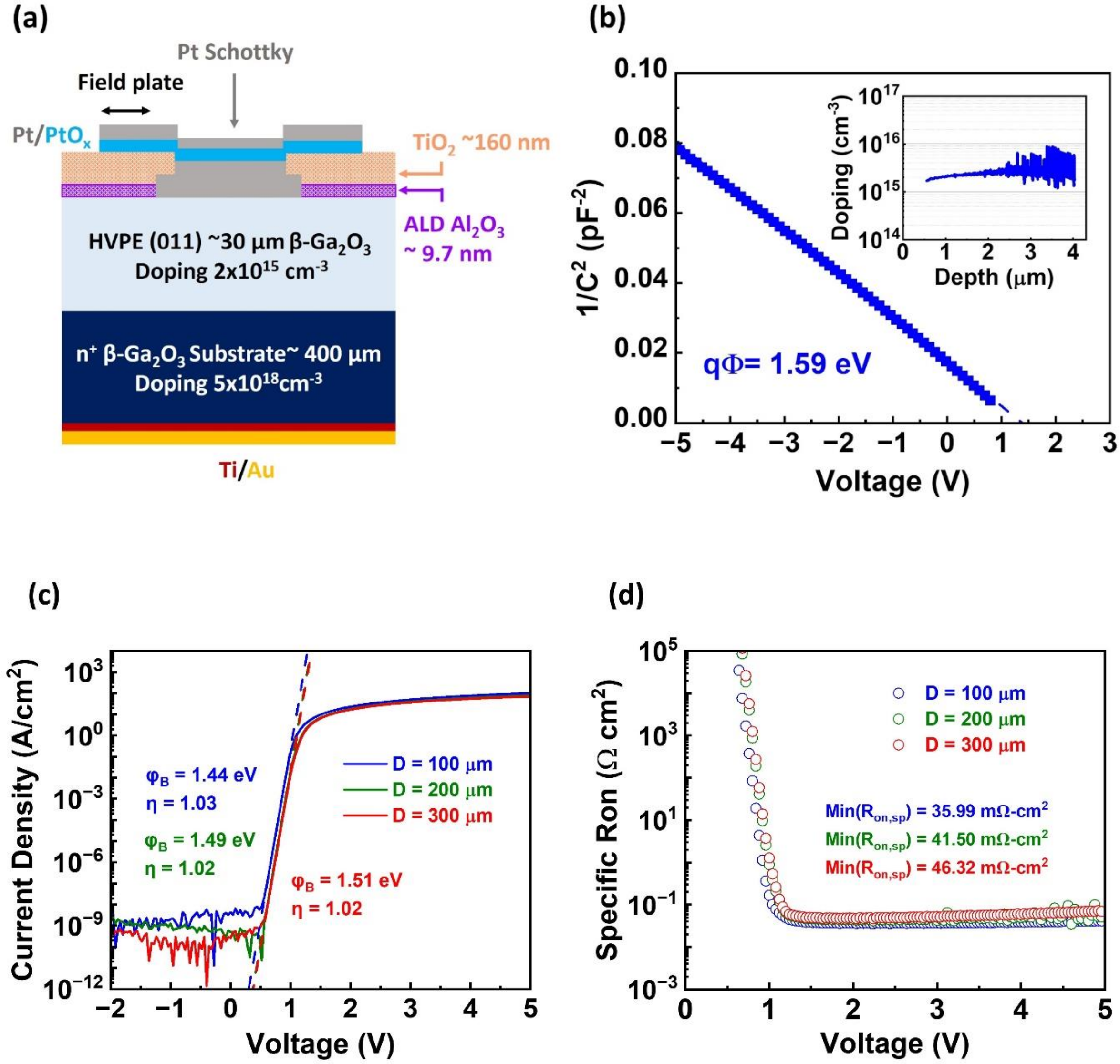

**Fig. 1.** (a) Schematic of the vertical (011) β-$Ga_2O_3$ Schottky diode with a 30 μm thick drift layer, Pt Schottky contact, and $TiO_2$/$Al_2O_3$ dielectric field-plate. (b) Extracted Schottky barrier height ($q\varphi_B$) of 1.59 eV from $1/C^2$-V analysis with the inset depicting C-V extracted doping concentrations. (c) Forward transport characteristics for the 100 μm, 200 μm, and 300 μm SBDs with field-plate exhibiting near-unity ideality factor (η) for all cases, and (d) the corresponding differential specific on-resistance ($R_{on,sp}$) and extracted minimum $R_{on,sp}$ for each case.

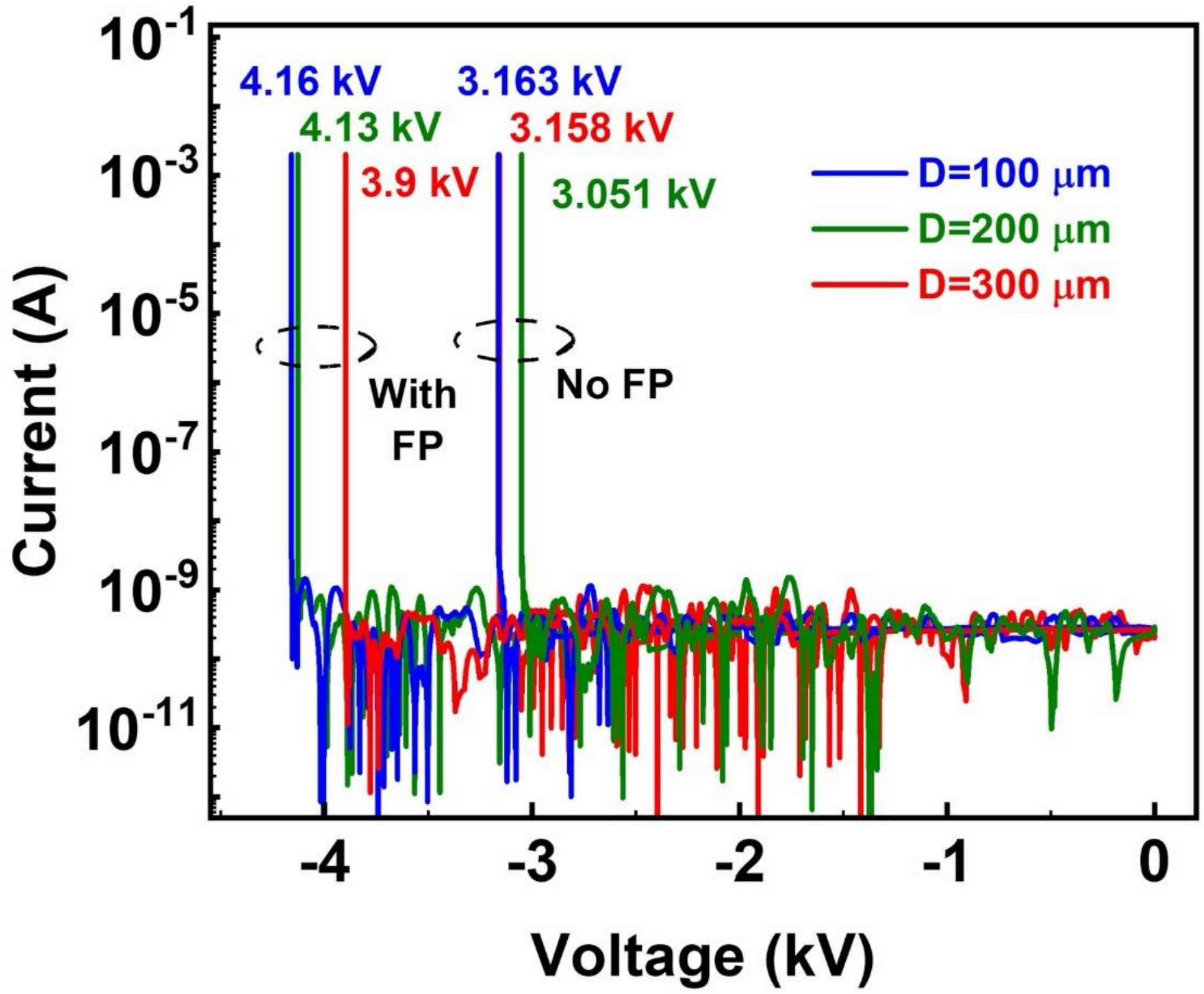

**Fig. 2.** Reverse current voltage (I-V) characterization for 100 μm, 200 μm, and 300 μm (011) β-$Ga_2O_3$ SBDs both with and without a field-plate. Devices showed similar breakdown voltages between 3.05 - 3.16 kV without a field-plate, and 3.90 - 4.16 kV with the inclusion of a dielectric field-plate. Devices maintained ultra-low leakage current prior to breakdown and showed similar breakdown voltages, which appear independent of diameter.

To understand the breakdown mechanism of the SBDs, we performed 2D Silvaco Atlas simulation of the 100 μm Pt SBD with the $TiO_2$/$Al_2O_3$ dielectric field-plate at its breakdown voltage $V_{br}$ = -4159 V [Fig. 3(a)]. The Silvaco simulation revealed a punch-through electric field profile with a parallel plane peak electric field of 1.9 MV/cm [Fig. 3(b)]. On the other hand, the electric field at the field-plate edge reached 4.3 MV/cm at the $Al_2O_3$/β-$Ga_2O_3$ interface while the peak electric field in the $TiO_2$ film appeared as 2.1 MV/cm [Fig. 3(b)]. It is to be mentioned that we also investigated the forward breakdown property of each dielectric component using vertical metal-oxide-semiconductor (MOS) diodes formed with β-$Ga_2O_3$. The ALD $Al_2O_3$ revealed a dielectric breakdown field of 6.2 MV/cm whereas the sputtered $TiO_2$ film showed degradation from high leakage current at 1.2 MV/cm. This indicates that the field-plate device breakdown was likely dictated by the dielectric failure of $TiO_2$.

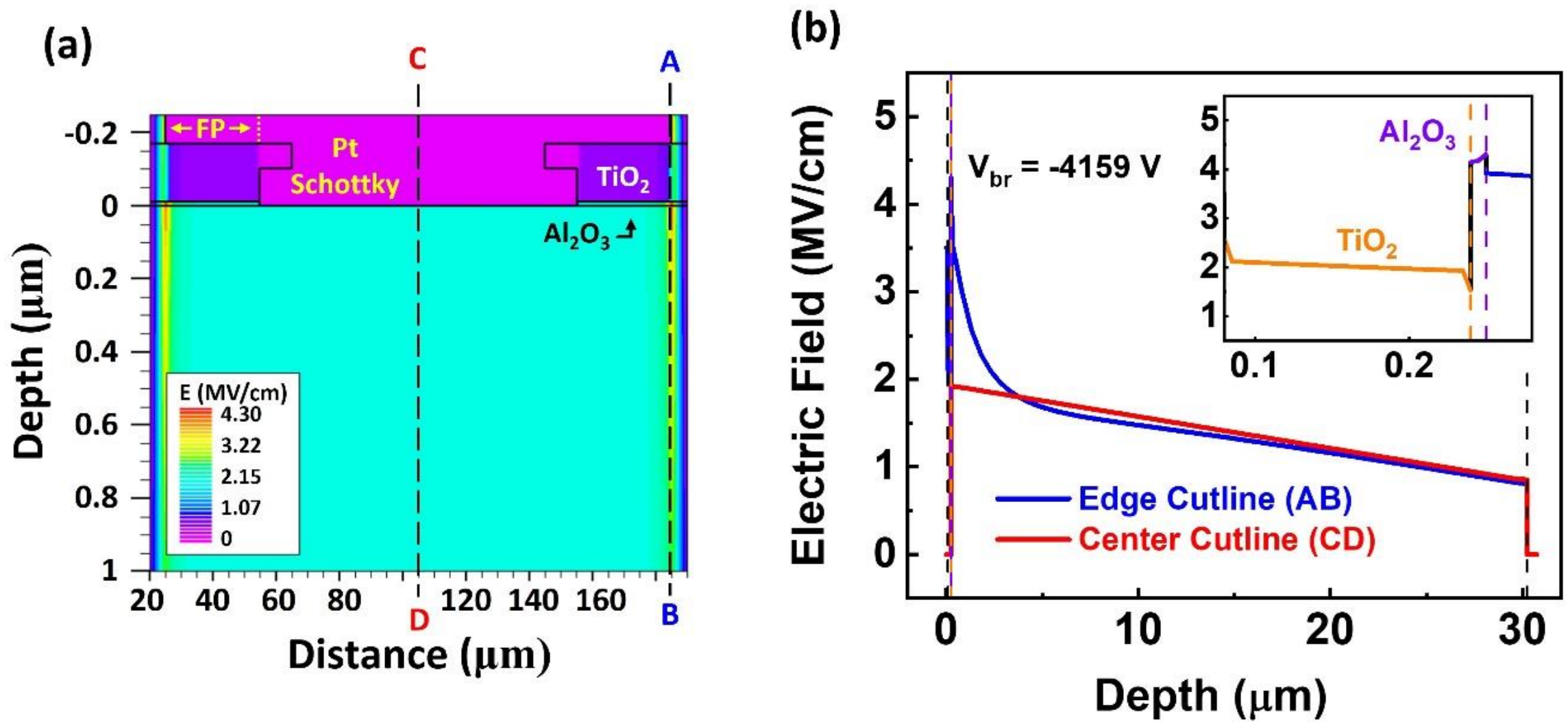

**Fig. 3.** (a) Simulated electric field contour at $V_{br}$ = -4159 V of the 100 μm diameter Pt Schottky diode with 30 μm field-plate using $TiO_2$/$Al_2O_3$ dielectric stack. (b) Electric field at field-plate edge extracted along cutline C-D showing a peak electric field of 4.3 MV/cm in $Al_2O_3$ while the $TiO_2$ dielectric contained an electric field of ~2.1 MV/cm (inset). The parallel plane electric field across Schottky center was extracted using cutline A-B that revealed a maximum parallel plane electric field of 1.9 MV/cm.

Subsequently, we investigated the high-temperature performance of the 100 μm diameter field-plate SBDs at both forward and reverse bias as a function of increased temperature from 25 °C to 150 °C (Fig. 4). The SBDs maintained an excellent rectification ratio ~$10^9$ throughout this temperature range [Fig. 4(a)]. The SBDs also exhibited near-unity ideality factor and reduced SBH at elevated temperatures [Fig. 4(c)], which is consistent with the expectation of an ideal thermionic emission transport.[31,38] We further investigated the temperature-dependent effects of the on-resistance ($R_{on}$). The forward J-V characteristics revealed an increase of $R_{on}$ at elevated temperature [Fig. 4(b)], which is primarily due to degradation of carrier mobility as reported in earlier works.[9,39] Using the power law, $R_{on}\,(T) = R_{on}^{300K}\,(\mathrm{T}/300)^{\alpha}$, the temperature coefficient of on-resistance (α) was extracted as 1.97 [Fig. 4(d)] which is comparable to the existing reports of β-$Ga_2O_3$ power diodes.[39,40]

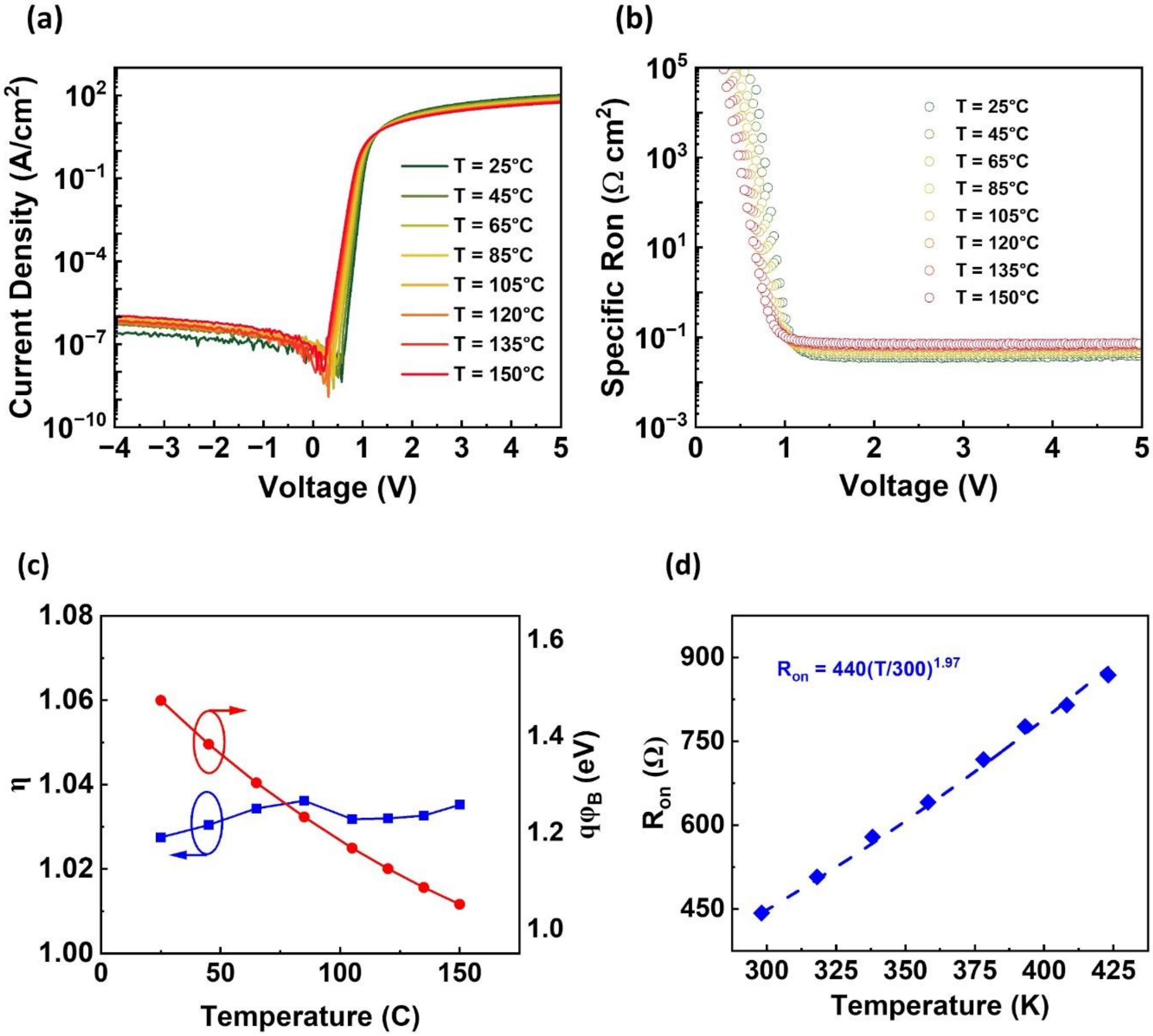

**Fig. 4.** (a) Temperature-dependent forward J-V characteristics for 100 μm field-plate (011) β-$Ga_2O_3$ SBD between 25 °C - 150 °C, (b) the corresponding differential specific on-resistance ($R_{on,sp}$). (c) the temperature-dependent SBH ($q\varphi_B$) and ideality factor (η). (d) On-resistance ($R_{on}$) dependence on temperature revealing the temperature coefficient α = 1.97 for the SBD.

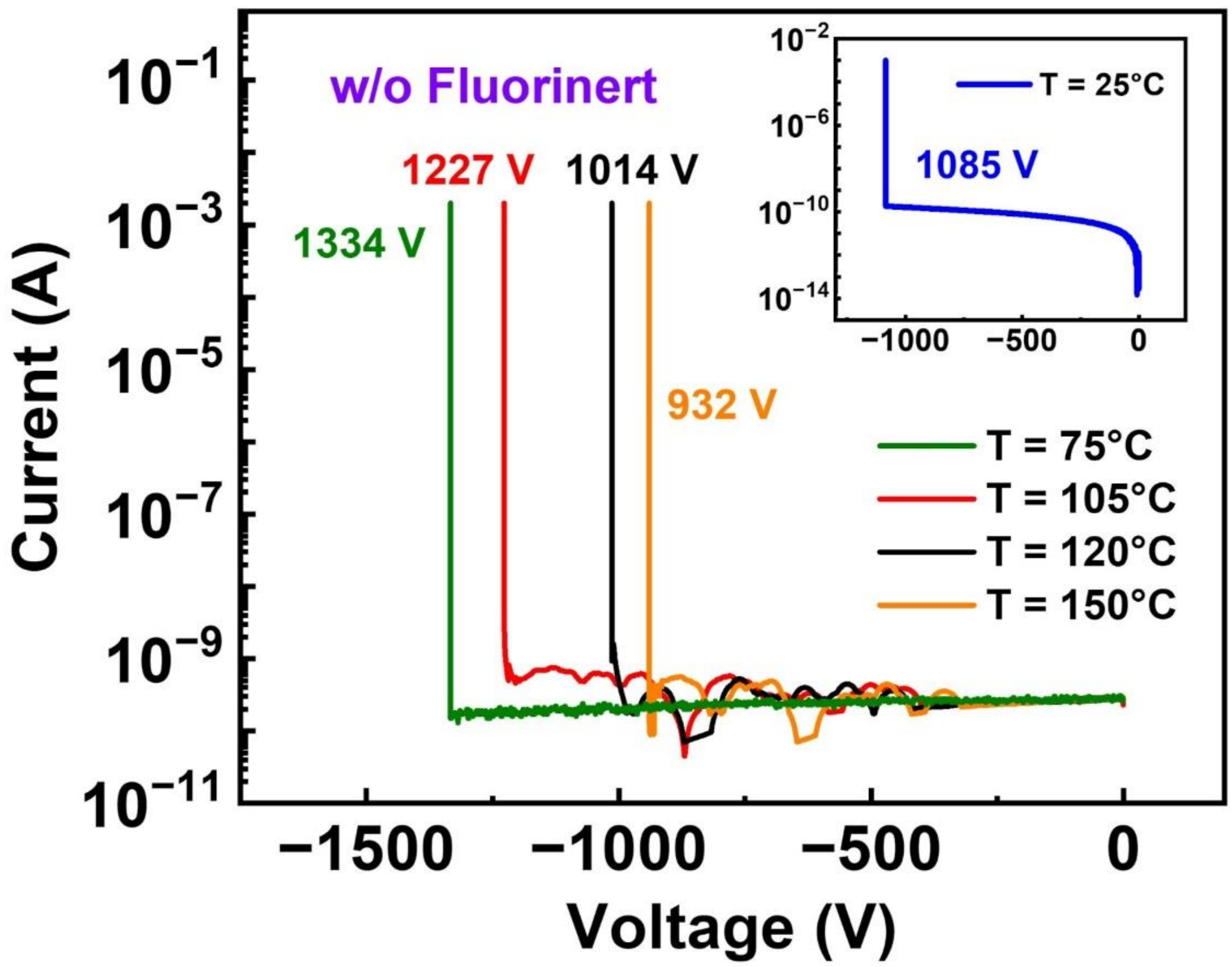

**Fig. 5.** Temperature-dependent reverse I-V characteristics of 100 μm (011) β-$Ga_2O_3$ field-plate SBD, tested from 75 °C to 150 °C, without Fluorinert submersion at AFRL probe station. The devices maintained similar low leakage current at high temperatures. The breakdown voltage of the devices was obtained in the range of ~1.0 - 1.33 kV, which was essentially limited by the air-breakdown due to the inability of using Fluorinert at high temperatures. The inset depicts a reference 100 μm diameter field-plate diode that was also tested without Fluorinert at room-temperature using the parametric analyzer at ISU. The room-temperature diode revealed a similar air-induced breakdown voltage of 1.08 kV.

We also explored temperature-dependent reverse breakdown performance using the 100 μm diameter field-plate SBDs up to 150 °C [Fig. 5]. Due to the high-temperature conditions, the reverse bias characterization was conducted without submerging the devices in Fluorinert. As the absence of Fluorinert can limit the device performances with premature air breakdown, we also measured the breakdown voltage of a 100 μm diameter field-plate SBD at room-temperature without Fluorinert, that was used as a baseline reference to compare the high-temperature breakdown performance of the SBDs (Fig. 5 inset). With increased temperature (75 - 150 ℃), the SBDs still exhibited kV-range breakdown voltage of ~1.0 - 1.33 kV despite being limited by the air breakdown effects. The premature air-breakdown was further confirmed by the reference room-temperature diode performance that also exhibited comparable breakdown voltage of 1.085 kV without Flurointert (Fig. 5 inset). Despite the air-induced breakdown, these (011) β-$Ga_2O_3$ SBDs maintained low leakage at elevated temperatures compared to the prior reports on vertical (001) β-$Ga_2O_3$ SBDs that suffered from orders of magnitude higher leakage current within this increased temperature range, which was attributed to trap-assisted effects.[7,10,31-33,38] Hence, our work demonstrates the strong potential of kV-class vertical (011) β-$Ga_2O_3$ SBDs to offer superior reverse blocking performance and minimal off-state leakage in high-temperature power devices owing to the reduced killer dislocation and low background impurity effects.

In summary, we demonstrated high-power vertical (011) β-$Ga_2O_3$ Schottky diodes fabricated with a high-permittivity $TiO_2/Al_2O_3$ dielectric field-plate that enabled excellent ~4 kV breakdown voltage with low leakage. Notably, this reverse blocking performance was similarly evident across different diode areas (100 - 300 µm) which shows great potential of the vertical (011) β-$Ga_2O_3$ SBDs towards scaling to large-area power devices. This outstanding device performance of the field-plate vertical (011) β-$Ga_2O_3$ was also obtained at high temperature, up to the explored 150 °C, where the diodes demonstrated excellent forward transport characteristics and kV-range reverse blocking ability while maintaining low leakage. As opposed to the conventional (001) β-$Ga_2O_3$ Schottky diodes that often suffer from temperature-dependent leakage and trap-assisted effects at high-voltage, the vertical (011) β-$Ga_2O_3$ SBDs did not exhibit any noticeable degradation of off-state current under the high-temperature and high-voltage condition. Hence, this work highlights the importance of combining advantageous material properties and improved field-management strategies for vertical (011) β-$Ga_2O_3$ β-$Ga_2O_3$ Schottky diodes to achieve superior performance in kV-class and high-temperature power switching applications.

**Acknowledgements**

This work was supported in part by NSF ECCS (Award No. 2401579) and AFOSR DEPSCOR program (Award No. FA9550-25-1-0292). A portion of this work was done in the UCSB Nanofabrication Facility, an open-access laboratory.

**Author Declarations**

The authors have no conflicts to disclose.

**Data Availability**

The data that support the findings of this study are available from the corresponding author upon reasonable request.

## References

[1] E. Farzana, S. Roy, N.S. Hendricks, S. Krishnamoorthy, and J.S. Speck, *Appl. Phys. Lett.* **123**(19), 192102 (2023).

[2] E. Ahmadi and Y. Oshima, *J. Appl. Phys.* **126**(16), 160901 (2019).

[3] E. J. Hollar and E. Farzana, *Appl. Phys. Lett.* **128**(5), 053503 (2026).

[4] A. M. Audri, C.-P. Ho, E. J. Hollar, J. Shi, and E. Farzana, *APL Electron. Devices* **1**(4), 046116 (2025).

[5] P. Adhikari, M. Ghassemi, E. J. Hollar, and E. Farzana, in *2026 IEEE Electr. Insul. Conf. EIC*, (2026).

[6] S. Islam, A. S. Senarath, E. Farzana, D. R. Ball, A. Sengupta, N. S. Hendricks, A. Bhattacharyya, R. A. Reed, E. X. Zhang, J. S. Speck, D. M. Fleetwood, and R. D. Schrimpf, *IEEE Trans. Nucl. Sci.* **71**(4), 515 (2024).

[7] K. Konishi, K. Goto, H. Murakami, Y. Kumagai, A. Kuramata, S. Yamakoshi, and M. Higashiwaki, *Appl. Phys. Lett.* **110**(10), 103506 (2017).

[8] E. Farzana, A. Bhattacharyya, N. S. Hendricks, T. Itoh, S. Krishnamoorthy, and J. S. Speck, *APL Mater.* **10**(11), 111104 (2022).

[9] B. Wang, M. Xiao, X. Yan, H. Yung Wong, J. Ma, K. Sasaki, H. Wang, and Y. Zhang, *Appl. Phys. Lett.* **115**, 263503 (2019).

[10] Z. Jian, S. Mohanty, and E. Ahmadi, *Appl. Phys. Lett.* **116**(15), 152104 (2020).

[11] F. Otsuka, H. Miyamoto, A. Takatsuka, S. Kunori, K. Sasaki, and A. Kuramata, *Appl. Phys. Express* **15**, 016501 (2022).

[12] W. Li, Z. Hu, K. Nomoto, Z. Zhang, J. Hsu, Q. T. Thieu, K. Sasaki, A. Kuramata, D. Jena, and H. G. Xing, *Appl. Phys. Lett.* **113**, 202101 (2018).

[13] P. Dong, J. Zhang, Q. Yan, Z. Liu, P. Ma, H. Zhou, and Y. Hao, *IEEE Electron Device Lett.* **43**(5), 765 (2022).

[14] S. Roy, B. Kostroun, J. Cooke, Y. Liu, A. Bhattacharyya, C. Peterson, B. Sensale-Rodriguez, and S. Krishnamoorthy*, Appl. Phys. Lett.* **123**, 243502 (2023).

[15] J. Williams, W. Wang, N. S. Hendricks, A. Adams, J. Piel, D. M. Dryden, K. Liddy, N. Sepelak, B. Morell, A. Islam, and A. Green, *J. Vac. Sci. Technol. A* **42**(3), 033405 (2024).

[16] C. Lin, Y. Yuda, M. H. Wong, M. Sato, N. Takekawa, K. Konishi, T. Watahiki, M. Yamamuka, H. Murakami, Y. Kumagai, and M. Higashiwaki, *IEEE Electron Device Lett.* **40**(9), 1487 (2019).

[17] S. Dhara, N. K. Kalarickal, A. Dheenan, C. Joishi, and S. Rajan, *Apply. Phys. Lett*. **121**, 203501 (2022).

[18] Z. Han, G. Jian, X. Zhou, Q. He, W. Hao, J. Liu, B. Li, H. Huang, Q. Li, X. Zhao, G. Xu, and S. Long, *IEEE Electron Device Lett.* **44**(10), 1680 (2023).

[19] N. Sun, H. H. Gong, T. C. Hu, F. Zhou, Z. P. Wang, X. X. Yu, F. Ren, S. L. Gu, H. Lu, R. Zhang, and J. D. Ye, *Appl. Phys. Lett.* **125**, 172104 (2024).

[20] W. Li, K. Nomoto, Z. Hu, D. Jena, and H. G. Xing, *IEEE Electron Device Lett.* **41**(1), 107 (2020).

[21] Z. Hu, Y. Lv, C. Zhao, Q. Feng, Z. Feng, K. Dang, X. Tian, Y. Zhang, J. Ning, H. Zhou, X. Kang, J. Zhang, and Y. Hao, *IEEE Electron Device Lett.* **41**(3), 441 (2020).

[22] T. Oshima and Y. Oshima, *Jpn. J. Appl. Phys.* **65**(13), 138001 (2026).

[23] D. Wakimoto, C.-H. Lin, K. Ema, Y. Ueda, H. Miyamoto, K. Sasaki, and A. Kuramata, *Appl. Phys. Express* **18**(10), 106502 (2025).

[24] S. Sdoeung, Y. Otsubo, K. Sasaki, A. Kuramata, and M. Kasu, *Appl. Phys. Lett.* **123**(12), 122101 (2023).

[25] B. Chen, W. Mu, Y. Liu, P. Wang, X. Ma, J. Zhang, X. Dong, Y. Li, Z. Jia, and X. Tao, *Cryst. Eng. Comm.* **25**(16), 2404 (2023).

[26] K. Goto, H. Murakami, A. Kuramata, S. Yamakoshi, M. Higashiwaki, and Y. Kumagai, *Appl. Phys. Lett.* **120**(10), 102102 (2022).

[27] M. M. Hossain Sarkar, D.S. Yu, J. Liu, Y. Ueda, C.-H. Lin, K. Sasaki, U. Singisetti, and H. Zhao, *Crystal Growth & Design*, (2026). https://doi.org/10.1021/acs.cgd.6c00857

[28] M. M. Hossain Sarkar, D. S. Yu, S. Alam, M. Hassan, J. Hwang, and H. Zhao, *ACS Appl. Electron. Mater.* **8**(3), 1380 (2026).

[29] J. Wen, J. Liu, Q. Li, L. Wang, Y. Ding, S. He, X. Zhou, S. Yang, G. Xu, and S. Long, in *2026 IEEE 38th Int. Symp. Power Semicond. Devices ICs ISPSD*, (2026), pp. 317.

[30] Y. Liu, C. Peterson, C. N. Saha, M. J. Tadjer, and S. Krishnamoorthy, "VBr >10 kV E-Beam/Sputtered Vertical NiOx/(011) β-Ga2O3 HJDs with PFOM >2.3 GW/cm$^2$," e-print arXiv:2604.27262 [physics.app-ph] (v1) (2026).

[31] M. Higashiwaki, K. Konishi, K. Sasaki, K. Goto, K. Nomura, Q.T. Thieu, R. Togashi, H. Murakami, Y. Kumagai, B. Monemar, A. Koukitu, A. Kuramata, and S. Yamakoshi, *Appl. Phys. Lett.* **108**(13), 133503 (2016).

[32] W. Hao, F. Wu, W. Li, G. Xu, X. Xie, K. Zhou, W. Guo, X. Zhou, Q. He, X. Zhao, S. Yang, and S. Long, *IEEE Trans. Electron Devices* **70**(4), 2129 (2023).

[33] H. H. Gong, X. X. Yu, Y. Xu, X. H. Chen, Y. Kuang, Y. J. Lv, Y. Yang, F. Ren, Z. H. Feng, S. L. Gu, Y. D. Zheng, R. Zhang, and J. D. Ye, *Appl. Phys. Lett.* **118**(20), 202102 (2021).

[34] Z. Hu, J. Li, C. Zhao, Z. Feng, X. Tian, Y. Zhang, Y. Zhang, J. Ning, H. Zhou, C. Zhang, Y. Lv, X. Kang, H. Feng, Q. Feng, J. Zhang, and Y. Hao, *IEEE Trans. Electron Devices* **67**(12), 5628 (2020).

[35] N. S. Hendricks, E. Farzana, A. E. Islam, K. D. Leedy, K. J. Liddy, J. Williams, D. M. Dryden, A. M. Adams, J. S. Speck, K. D. Chabak, and A. J. Green, *Appl. Phys. Express* **16**(7), 071002 (2023).

[36] T.-H. Hung, K. Sasaki, A. Kuramata, D. N. Nath, P. Sung Park, C. Polchinski, and S. Rajan, *Appl. Phys. Lett.* **104**(16), 162106 (2014).

[37] E. Farzana, Z. Zhang, P. K. Paul, A. R. Arehart, and S. A. Ringel, *Appl. Phys. Lett.* **110**(20), 202102 (2017).

[38] C. Fares, F. Ren, and S. Pearton, *ECS J. Solid State Sci. Technol*, **8 (7),** Q3007 (2019).

[39] C. Buttay, H. Wong, B. Wang, M. Xiao, C. Dimarino, and Y. Zhang, *Microelectron. Reliab.* **114**, 113743 (2020).

[40] Y. Liu, S. Roy, C. Peterson, A. Bhattacharyya, and S. Krishnamoorthy, *AIP Adv.* **15**(1), 015114 (2025).